\documentclass{iaus}
\usepackage{graphicx}

\newcommand{\apj}{{\it ApJ}}
\newcommand{\aj}{{\it AJ}}
\newcommand{\mnras}{{\it MNRAS}}
\newcommand{\aanda}{{\it A\&A}}
\newcommand{\pasp}{{\it PASP}}

\title[IAUS289.~~Cepheid distances from the Baade--Wesselink method] 
{Cepheid distances from the Baade--Wesselink method}

\author[Gieren et al.]   
{Wolfgang Gieren,$^1$ Jesper Storm,$^2$ Nicolas Nardetto,$^3$\\
 Alexandre Gallenne,$^1$ Grzegorz Pietrzy\'nski,$^1$ Pascal
 Fouqu\'e,$^4$\\ Thomas G. Barnes,$^5$ \and Daniel Majaess$^6$}

\affiliation{
$^1$Universidad de Concepci\'on, Departamento de
  Astronom\'{\i}a,Casilla 160-C, Concepci\'on, Chile\\ email: {\tt
    wgieren@astro-udec.cl} \\[\affilskip]
$^2$Leibniz-Institut f\"ur Astrophysik Potsdam (AIP), An der
  Sternwarte 16, 14482 Potsdam, Germany\\ email: {\tt jstorm@aip.de}
  \\[\affilskip] 
$^3$Laboratoire Lagrange, UMR7293, UNS/CNRS/OCA, 06300 Nice,
  France\\[\affilskip]
$^4$IRAP, Universit\'e de Toulouse, CNRS, 14 Avenue E. Belin, 31400
  Toulouse, France\\[\affilskip]
$^5$University of Texas at Austin, McDonald Observatory, 82 Mt. Locke
  Road, McDonald Observatory, TX 79734, USA\\[\affilskip]
$^6$Department of Astronomy and Physics, Saint Mary's University,
  Halifax, NS B3H 3C3, Canada\\[\affilskip]
}

\pubyear{2012}
\volume{289}  
\jname{Advancing the Physics of Cosmic Distances}
\editors{Richard de Grijs \& Giuseppe Bono, eds.}
\begin{document}

\maketitle

\begin{abstract}
Recent progress on Baade--Wesselink (BW)-type techniques to determine
the distances to classical Cepheids is reviewed. Particular emphasis
is placed on the near-infrared surface-brightness (IRSB) version of
the BW method. Its most recent calibration is described and shown to
be capable of yielding individual Cepheid distances accurate to 6\%,
including systematic uncertainties. Cepheid distances from the IRSB
method are compared to those determined from open cluster zero-age
main-sequence fitting for Cepheids located in Galactic open clusters,
yielding excellent agreement between the IRSB and cluster Cepheid
distance scales. Results for the Cepheid period--luminosity (PL)
relation in near-infrared and optical bands based on IRSB distances
and the question of the universality of the Cepheid PL relation are
discussed. Results from other implementations of the BW method are
compared to the IRSB distance scale and possible reasons for
discrepancies are identified.
\end{abstract}




\firstsection 
\section{Introduction}

Since its original conception by \cite[Baade (1926)]{Baade_1926_11_0}
and \cite[Wesselink (1946)]{Wesselink_1946_01_0}, the
Baade--Wesse\-link (BW) method has been used, and refined over time,
to calculate the radii of and distances to radially pulsating stars,
in particular classical Cepheid variables. The method fits the
observed radial velocity curves of Cepheids to their angular-diameter
curves (derived interferometrically or by some equivalent means). This
allows to establish an independent astrophysical distance scale, which
is extremely useful to check Cepheid distances derived using other
methods, particularly based on Cepheids in open clusters, and to
independently establish period--luminosity (PL) relations in both the
Milky Way and the Magellanic Clouds (e.g., \cite[Storm
  et~al. 2011a,b]{Storm_2011_10_0,Storm_2011_10_1}). In this paper, we
review the state of the art in BW distance determination, focusing on
using the most recent calibrations of the method reported in the
literature.

\section{Distances based on the infrared surface-brightness (IRSB) version of the Baade--Wesselink method}

\begin{figure}[!h]
\begin{center}
 \includegraphics[width=4in]{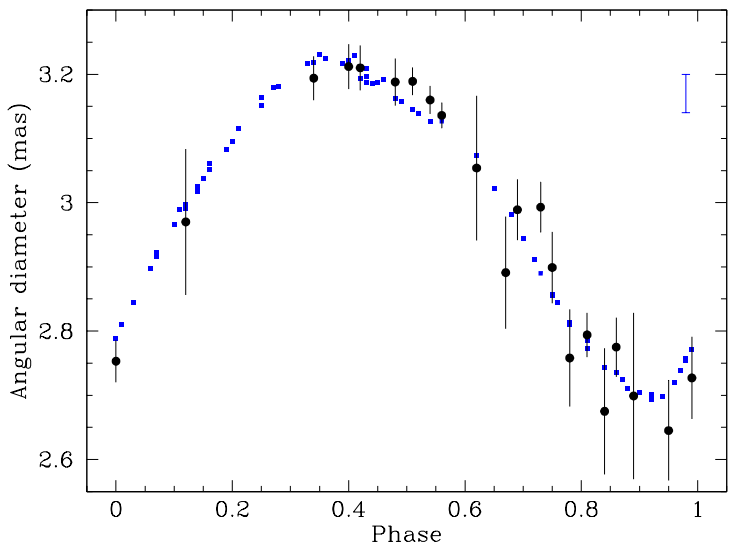} 
 \caption{Solid bullets: Interferometric angular-diameter measurements
   for the Cepheid $\ell$ Car. Blue squares: Angular diameters
   calculated using the IRSB technique. The agreement is at the 1\%
   level (see \cite[Kervella
   et~al. 2004b]{Kervella_2004_04_0}).}  \label{gierfig1}
\end{center}
\end{figure}

The infrared surface-brightness (IRSB) version of the BW method was
introduced by \cite[Fouqu\'e \& Gieren (1997)]{Fouque_1997_04_0}. It
employs the $(V-K)$ color of Cepheid variables to calculate their
surface brightnesses, from which the angular diameters at the
corresponding pulsation phases are derived. In this way, $VK$
photometry of a Cepheid through its pulsation cycle provides its
angular-diameter curve. In the original calibration of the
surface-brightness--color relation of \cite[Fouqu\'e \& Gieren
  (1997)]{Fouque_1997_04_0}, interferometrically determined angular
diameters of non-pulsating giants and supergiants with colors similar
to Cepheids were used. The surface-brightness--color relation later
obtained from phase-resolved angular diameters of a sample of nearby
Cepheids by \cite[Kervella et~al. (2004a)]{Kervella_2004_12_0} with
the European Southern Observatory's (ESO) {\sl Very Large Telescope
  Interferometer} demonstrated that the relation, obtained directly
from the Cepheids, agrees to better than 2\% with that derived from
stable giants \cite[(Kervella et~al. 2004b)]{Kervella_2004_04_0}. An
example is shown in Fig.\,\ref{gierfig1} for $\ell$~Car, the Cepheid with
the largest angular diameter in the sky.

To calculate the radius variation of a Cepheid from its observed
radial velocity curve, the $p$ factor---which converts a radial to a
pulsational velocity---must be known for the star of interest. In
earlier work (\cite[Gieren et al. 1993, 1997, 1998; Storm et
  al. 2004]{Gieren_1993_11_0,Gieren_1997_10_0,Gieren_1998_03_0,Storm_2004_02_0}),
a slightly period-dependent $p$ factor was adopted based on the models
of \cite[Hindsley \& Bell (1986)]{Hindsley_1986_09_0}. In \cite[Gieren
  et al. (2005)]{Gieren_2005_07_0} the IRSB technique was applied for
the first time to a small sample (13 stars) of Large Magellanic Cloud
(LMC) Cepheids, producing an unexpectedly significant and unphysical
period dependence of the distance moduli of LMC Cepheids. This
suggested a stronger dependence of the $p$ factor on pulsation period
than predicted by \cite[Hindsley \& Bell (1986)]{Hindsley_1986_09_0}
and led to the relation $p = 1.58 - 0.15 \log P$ (days). \cite[Storm
  et~al. (2011a,b)]{Storm_2011_10_0,Storm_2011_10_1} confirmed the
\cite[Gieren et al. (2005)]{Gieren_2005_07_0} result of a steep
$p$-factor law by extending the analysis to 36 Cepheids in the LMC for
which very accurate radial velocity curves were obtained with the
HARPS instrument at ESO/La Silla, and for which accurate photometry
was available from \cite[Persson et al. (2004)]{Persson_2004_11_0} in
the $K$ band and from {\sc ogle iii} in the $V$ band. Simultaneously,
\cite[Storm et~al. (2011a,b)]{Storm_2011_10_0,Storm_2011_10_1} also
obtained new radial velocity data for a number of Milky Way (MW)
Cepheids. Based on the combined MW and LMC Cepheid data set, the IRSB
technique was re-calibrated using the two constraints that (i) the
IRSB Cepheid distances must agree with the {\sl Hubble Space
  Telescope} parallax distances of 9 MW Cepheids (\cite[Benedict et
  al. 2007]{Benedict_2007_04_0}; the binary Cepheid W Sgr was excluded
because its bright blue companion produces a spurious IRSB solution),
and (ii) that the tilt-corrected (\cite[van der Marel \& Cioni
  2001]{van-der-Marel_2001_10_0}) LMC Cepheid distances do not depend
on their pulsation periods. These requirements led to a $p$-factor
law, $p = 1.55 -0.186 \log P$ (days), which agrees within the
uncertainties with the \cite[Gieren et al. (2005)]{Gieren_2005_07_0}
result.

\begin{figure}[!h]
\begin{center}
 \includegraphics[width=4in]{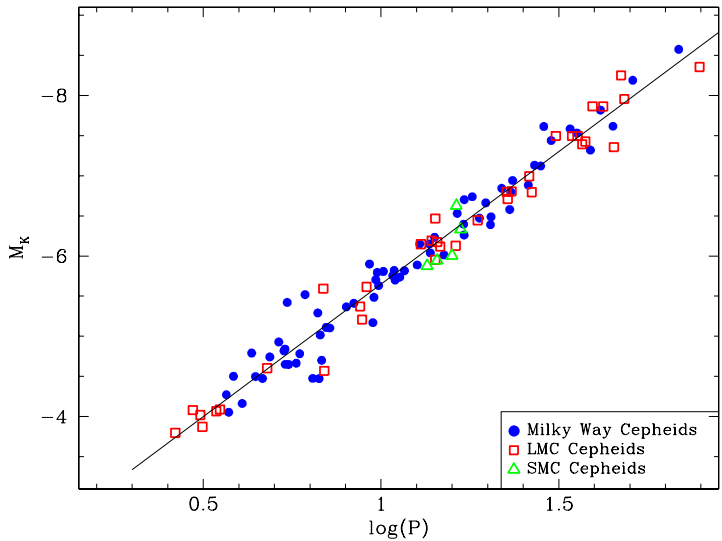} 
 \caption{Composite $K$-band Cepheid PL relation obtained from IRSB
   distances to 70 Cepheids in the MW, 36 in the LMC, and five in the
   Small Magellanic Cloud (SMC). The $K$-band PL relation is
   independent of metallicity in the range from solar to the typical
   SMC Cepheid metallicity.}
   \label{gierfig2}
\end{center}
\end{figure}

\cite[Storm et~al. (2011a,b)]{Storm_2011_10_0,Storm_2011_10_1} used
their new calibration of the IRSB method to provide PL relations for
the MW and LMC Cepheid samples in the optical $V, I,$ and ($V-I$)
Wesenheit bands and in the near-infrared $J$ and $K$
bands. Particularly in the near-infrared bands, the Galactic and LMC
PL relations show excellent agreement for both their slopes and zero
points. In the $K$ band, the relations are $M_K = (-3.33 \pm 0.09)
(\log P - 1) - (5.66 \pm 0.03)$ (MW), and $M_K = (-3.28 \pm 0.09)
(\log P - 1) - (5.64 \pm 0.04)$ (LMC). There is also excellent
agreement with the slope observed by \cite[Persson et
  al. (2004)]{Persson_2004_11_0} for the LMC $K$-band PL relation
($-3.26 \pm 0.02$). These results suggest that the $K$-band PL
relation for classical Cepheids is universal across the metallicity
range from solar down to the typical LMC Cepheid metallicity
(approximately $-0.4$\,dex; \cite[Luck et al. 1998]{Luck_1998_02_0};
\cite[Romaniello et al. 2008]{Romaniello_2008_09_0}). The excellent
agreement of the MW and LMC PL relations at $K$ allows us to construct
a common PL diagram, which is shown in Fig.\,\ref{gierfig2} and can be
used directly for distance determination to galaxies containing
sizeable samples of classical Cepheids. The distance to the LMC's
barycenter implied by the Storm et al. results is $(m-M)_0 = 18.45 \pm
0.04$\,(statistical)\,mag. The MW and LMC PL relations in the optical
$V$ and $I$ bands agree within their uncertainties, but there is a
marginally significant metallicity effect in the $W_I$ band, $-0.23
\pm 0.10$\,mag dex$^{-1}$ (more metal-rich Cepheids are more luminous
for a given period than their more metal-poor counterparts).

\section{Comparison of IRSB and cluster zero-age main-sequence-fitting Cepheid distance scales}

\begin{figure}[!h]
\begin{center}
 \includegraphics[width=4in]{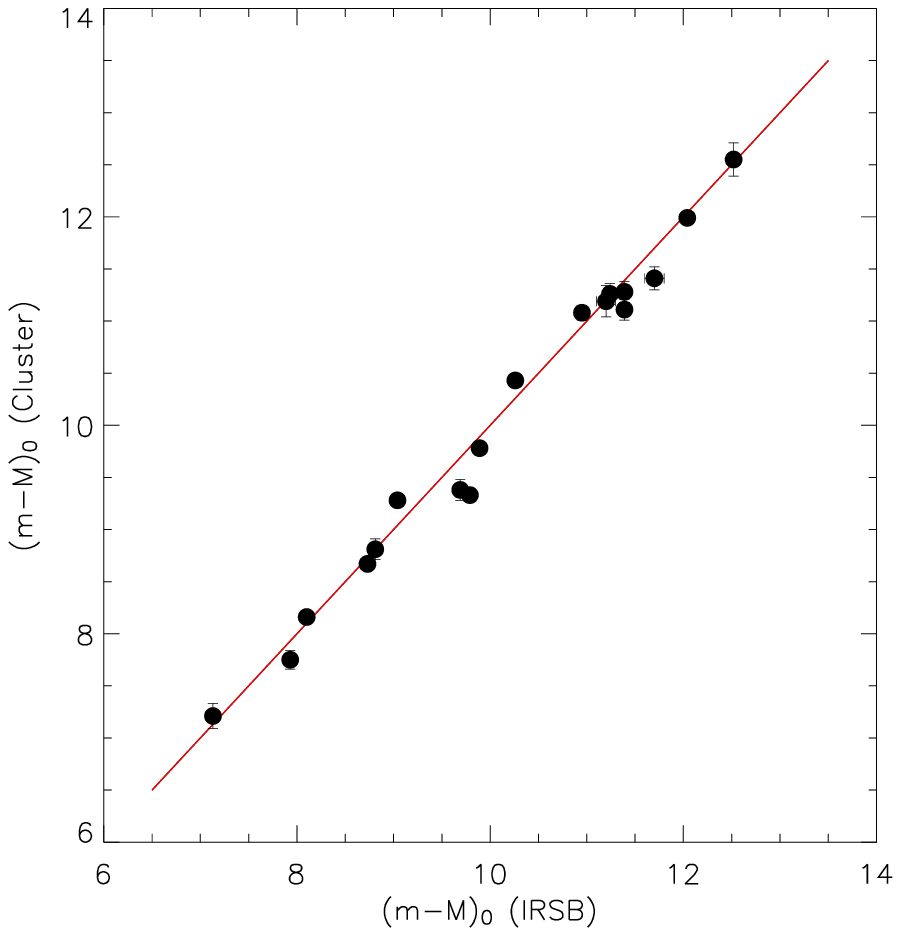} 
 \caption{ZAMS-fitting distance moduli for 18 open cluster Cepheids
   versus their IRSB distance moduli calculated based on the
   prescription of \cite[Storm et~al. (2011a)]{Storm_2011_10_0}. The
   agreement between the open cluster and IRSB distance scales is at
   the 3\% level.}
   \label{gierfig3}
\end{center}
\end{figure}

IRSB distances were calculated for 18 Galactic open cluster Cepheids
whose cluster membership was established with a high level of
confidence. Their zero-age main-sequence (ZAMS)-fitting distances were
adopted from \cite[Turner (2010)]{Turner_2010_04_0} and supplemented
with the recent ZAMS-fitting distances to the cluster Cepheids TW~Nor
(\cite[Majaess et al. 2011]{Majaess_2011_11_0}), SU~Cas (\cite[Turner
  et al. 2012]{Turner_2012_05_0}), $\delta$~Cep (\cite[Majaess et
  al. 2012a]{Majaess_2012_03_0}), and $\xi$~Gem (\cite[Majaess et
  al. 2012b]{Majaess_2012_03_1}). From these data, we find a mean
difference, $(m-M)_\mathrm{ZAMS} - (m-M)_\mathrm{IRSB} = +0.07 \pm
0.09$\,mag, showing that the IRSB and cluster Cepheid distance scales
currently agree at the 3\% level. This excellent agreement between the
two completely independent distance scales is exemplified by
Fig.\,\ref{gierfig3}.

\section{Results from other implementations of the Baade--Wesselink method}

Other recent implementations of the BW method, such as that used by
\cite[Groenewegen (2007)]{Groenewegen_2007_11_0} or \cite[Joner \&
  Laney (2012)]{Joner_2012_05_0}, have found (on average) rather
similar distances to individual Cepheids despite their use of
different $p$ factors for the conversion from radial to pulsational
velocities. \cite[Groenewegen (2007)]{Groenewegen_2007_11_0} finds
evidence of a constant $p$ factor of 1.27 (albeit with a large scatter
for individual Cepheids) based on his calibrating sample with adopted
distances from their known parallaxes. \cite[Joner \& Laney
  (2012)]{Joner_2012_05_0} adopt a slope of $-0.07$, which is in good
agreement with the theoretical prediction of \cite[Nardetto et
  al. (2009)]{Nardetto_2009_05_0}. Both versions of the BW technique
actually derive the mean radii of the Cepheids, which are then
transformed to luminosities and distances adopting a particular
effective temperature scale.

Very recently, the CORS method (\cite[Caccin et
  al. 1981]{Caccin_1981_04_0}) was re-calibrated by \cite[Molinaro et
  al. (2011)]{Molinaro_2011_05_0} and applied to the massive young LMC
cluster NGC 1866, which contains more than 20 classical Cepheid
variables (\cite[Molinaro et al. 2012]{Molinaro_2012_03_0}). The CORS
BW technique is a semi-theoretical approach; it uses Walraven
photometry and calibrates the Cepheid surface brightness using a grid
of atmosphere models. Adopting $p = 1.27$, the $V$-band magnitudes of
Molinaro et al. (2012) are systematically brighter, by approximately
0.2\,mag, than the $V$-band magnitudes of \cite[Storm et
  al. (2011a)]{Storm_2011_10_0}. The likely overestimation of Cepheid
absolute magnitudes, and, thus, their distances based on the CORS
method becomes evident in the application of the technique to five
Cepheids in NGC 1866 with periods clustering around 3\,days, that are
in common with the \cite[Storm et al. (2011b)]{Storm_2011_10_1}
sample. If Storm et al. had used $p = 1.27$ instead of $p = 1.46$ as
predicted by the IRSB $p$-factor law for Cepheids of $P \sim 3$\,days,
the IRSB distance modulus of NGC 1866 would become 18.14\,mag, which
is clearly unreasonably short, and $\sim 0.4$\,mag shorter than the
CORS mean distance for these five variables. On the other hand, the
CORS distance modulus to NGC 1866 is 18.57\,mag, which is only
0.12\,mag larger than the average LMC distance derived from the IRSB
method by \cite[Storm et
  al. (2011a,b)]{Storm_2011_10_0,Storm_2011_10_1}.

It is important to stress that the only version of the BW method which
has used LMC Cepheids as a constraint for the calibration (demanding
that LMC Cepheid distances must be independent of their periods) is
the IRSB technique as calibrated by \cite[Storm et al. (2011a,b)]{},
which places additional weight on this particular BW method.

\section{Discussion}

The most important current systematic uncertainty affecting any
version of the BW method is which $p$-factor law to adopt. Recently,
\cite[Ngeow et al. (2012)]{Ngeow_2012_07_0} used revised {\sl
  Hipparcos} parallaxes (9 Cepheids) and ZAMS-fitting distances (16
cluster Cepheids) for a sample of 25 MW Cepheids in common with
\cite[Storm et al. (2011a)]{Storm_2011_10_0} to derive their
individual $p$-factor values by requiring that the independent
distances should agree with the IRSB distance determination. This work
fully confirmed the steep $p$-factor law of Storm et al. A very recent
study by \cite[Groenewegen (2013)]{Groenewegen_2013_in_press} using MW
and LMC Cepheids also confirms the steep Storm et al. $p$-factor law,
resulting in an even steeper dependence of $p$ on the pulsation
period. This makes it very likely that the conclusions about the
universality of the Cepheid PL relation reached by Storm et al. are
correct. How is it then possible that other BW-method implementations
(perhaps except the latest implementation of the CORS method) using
different $p$-factor laws quite frequently produce distances to
individual Cepheids that are consistent with the IRSB determination?
The answer seems to be, at least in part, that different
implementations of the BW method require different $p$-factor
laws. This, in turn, tells us that there is physics hidden in the $p$
factor which is not yet fully understood or taken into account in the
theoretical modeling of Cepheid atmospheres. This may be the reason
why these models currently predict a relatively shallow dependence of
$p$ on period, in contrast to the IRSB results (\cite[Nardetto et
  al. 2009]{Nardetto_2009_05_0}; also this volume). It is also
possible (likely) that the $p$-factor relation for classical Cepheids
exhibits an intrinsic dispersion. It seems to be independent, however,
of the metallicity (\cite[Nardetto et al. 2011]{Nardetto_2011_10_0}).

Future precision measurements of the $p$ factors of Cepheids,
particularly of short-period Cepheids, will be very important. There
is hope that direct interferometric angular-diameter measurements
(e.g., \cite[M\'erand et al. 2005]{Merand_2005_07_0}; \cite[Gallenne
  et al. 2012]{Gallenne_2012_03_0}) will improve sufficiently to meet
that goal. An interesting alternative is the analysis of Cepheids in
eclipsing binaries, allowing an accurate distance determination based
on application of the binary method (\cite[Pietrzynski et al. 2009,
  2010,
  2011]{Pietrzynski_2009_05_0,Pietrzynski_2010_11_0,Pietrzynski_2011_12_0}),
which promises to yield accurate $p$-factor measurements for these
binary Cepheids.

Another possibly important source of systematic uncertainties in BW
distances derived from infrared photometry is the recent detection of
circumstellar shells around Cepheids (\cite[Kervella et al. 2006;
  M\'erand et al. 2006, 2007; Gallenne et
  al. 2011]{Kervella_2006_03_0,Merand_2006_07_0,Merand_2007_08_0,Gallenne_2011_11_0}). The
effect on the flux in the $K$ band is 1--4\% for the (few) Cepheids
which have been studied in this context so far, which may affect
(reduce) the IRSB distances by up to 4\%. It has yet to be seen how
ubiquitous such shells are in order to assess more reliably their
possible effect on BW distance determinations. If there are few
Cepheids with strong flux contributions from circumstellar shells,
these stars could be omitted from samples used for calibration or
distance determination, similarly to binary Cepheids which are known
members of systems hosting relatively bright companion stars.

Currently, BW distance determinations based on a variety of different
implementations of the method reach statistical uncertainties of
1--2\% for Cepheids using excellent existing data sets. The systematic
uncertainties affecting these distances limit current BW distances to
individual Cepheids to approximately 6\% accuracy. There is room for
improvement over the next few years along the lines discussed
here. Theoretical models still seem to miss important physics and must
be improved to produce predictions which can be reconciled with
observational facts.

\bigskip

\acknowledgements WG and GP gratefully acknowledge financial support
for this work from the BASAL Centro de Astrof\'{\i}sica y
Tecnolog\'{\i}as Afines (CATA) PFB-06/2007. Support from Polish grant
N203 387337 and a TEAM subsidy of the Foundation for Polish Science
(FNP) is also acknowledged.

\end{document}